\documentclass[conference]{IEEEtran}
\IEEEoverridecommandlockouts
\usepackage{cite}
\usepackage{amsmath,amssymb,amsfonts}
\usepackage{algorithmic}
\usepackage{graphicx}
\usepackage{url}
\usepackage{textcomp, cancel}
\usepackage{float}
\usepackage{stackengine}
\usepackage{subcaption}
\usepackage{xcolor}
\usepackage{tabulary}
\usepackage{multicol}% http://ctan.org/pkg/multicols
\usepackage{graphicx}% http://ctan.org/pkg/graphicx
\def\BibTeX{{\rm B\kern-.05em{\sc i\kern-.025em b}\kern-.08em
    T\kern-.1667em\lower.7ex\hbox{E}\kern-.125emX}}
    
\DeclareMathOperator{\EX}{\mathbb{E}}% expected value

\definecolor{celticGreen}{HTML}{008A3B}

\newcommand{\armar}{\IEEEauthorrefmark}
\DeclareMathOperator{\Tr}{Tr}

\begin{document}

% % %
% Title and Authorship
\title{Coupling Rendering and Generative Adversarial Networks for Artificial SAS Image Generation}

\makeatletter
\newcommand{\linebreakand}{%
  \end{@IEEEauthorhalign}
  \hfill\mbox{}\par
  \mbox{}\hfill\begin{@IEEEauthorhalign}
}
\makeatother
% https://tex.stackexchange.com/questions/458204/ieeetran-document-class-how-to-align-five-authors-properly
\author{
  \IEEEauthorblockN{Albert Reed\armar{1}, Isaac D. Gerg\armar{2}\armar{5}, John D. McKay\armar{2}, Daniel C. Brown\armar{2}\armar{3}, David P. Williams\armar{6}, Suren Jayasuriya\armar{1}\armar{4}}\hspace{0.3cm}
  \linebreakand
  \IEEEauthorblockA{\textit{\armar{1}School of Electrical, Computer} \\\textit{\& Energy Engineering,}\\\textit{\armar{4}School of Arts, Media \& Engineering,} \\
    \textit{Arizona State University}\\
    Tempe, AZ USA}
  \and
  \IEEEauthorblockA{\textit{\armar{2}Applied Research Laboratory,}\\\textit{ \armar{3}Graduate Program in Acoustics,} \\ \textit{\armar{5}School of Electrical Engineering} \\
  \textit{and Computer Science,}
  \\
    \textit{Pennsylvania State University}\\
    State College, PA USA}
  \and
  \IEEEauthorblockA{\textit{\armar{6}NATO STO Centre for Maritime}\\\textit{Research \& Experimentation,}\\
    La Spezia, Italy}
}
\maketitle

% % % 
% Introduction
\section{Introduction}

There is a growing demand for large-scale Synthetic Aperture Sonar (SAS)  datasets. This demand stems from data-driven applications such as Automatic Target Recognition (ATR)~\cite{stack2011automation, OnHumanPerception, underwater_target}, segmentation~\cite{mignotte2000sonar} and oceanographic research of the seafloor, simulation for sensor prototype development and calibration~\cite{blanford2018design}, and even potential higher level tasks such as motion estimation~\cite{doisy1998general} and micronavigation~\cite{bellettini2002theoretical}. Unfortunately, the acquisition of SAS data is bottlenecked by the costly deployment of SAS imaging systems, and even when data acquisition is possible, the data is often skewed towards containing barren seafloor rather than objects of interest. This skew introduces a data imbalance problem wherein a dataset can have as much as a 1000-to-1 ratio of seafloor background to object-of-interest SAS image chips. 

An alternative to real-world SAS image capture is to generate artificial SAS images that can be used to construct large-scale datasets. This has been approached through either model-driven, physics-based approaches or more recently through data-driven, machine learning approaches such as generative adversarial networks (GANs). These two methods have seen relative levels of success in synthesizing SAS data. Physics-based models, such as scattering models~\cite{Brown:2017b,swat}, allow for absolute user specification and control of environment and SAS physics interactions, and simulate physical realistic effects such as speckle and spatial coherence length. However, these models induce an intractable computational burden (i.e. hours for a single image), and are unable to produce SAS realistic images due to the complexity of modeling the entire scene and environment explicitly. On the other hand, data-driven models such as GANs can rapidly generate a large number of realistic SAS images that match input distribution statistics. However, the user has little control or specification over scene content in these images, and we will show that these models cannot generalize and generate truly novel images, especially when trained in the data-starved SAS regime. 

To leverage the strengths of both model  and data-driven approaches, we propose a hybrid pipeline combining the two. In particular, we utilize an optical renderer coupled with a GAN. The optical renderer serves as a method for quickly rendering an image that approximates the interactions of SAS systems with objects on the seafloor, and the GAN ingests these images as input and colors them with SAS-realistic properties which are traditionally hard to model in closed form. 

\textbf{Our specific contributions are as follows:}
\begin{enumerate}
\item  Use of an open source physically-based optical ray tracer to render artificial SAS images.
\item  A Wasserstein generative adversarial network to color these optically rendered SAS images with the visual and statistical qualities of real SAS images.
\item  A hybrid pipeline from optical rendering to a GAN that allows for control over target and scene geometry, background, and sources while maintaining this SAS image realism. 
\end{enumerate}
These contributions are validated on a real SAS dataset~\cite{williams2019novel}, and we perform qualitative and quantitative analysis to study both the benefits and drawbacks of our approach. We hope this work spurs more synergistic combinations of physics-based modeling with data-driven machine learning approaches for SAS in the future.

% % %
% Related Work
\section{Related Work}

% %
% Generative Adversarial Networks
\subsection{Generative Adversarial Networks}
\label{sec:GANrelated}

We introduce the literature surrounding generative adversarial networks (GANs) for those unfamiliar with these machine learning models. Generative adversarial networks (GANs), invented by Goodfellow et al.~\cite{goodfellow_GANs}, are generative models that learn to generate new data that follow the distribution statistics of a given training dataset. For example, a GAN trained on images of cats will produce images of cats not seen before in the training data. GANs have been used to generate photorealistic faces of fake celebrities~\cite{karras2017progressive}, new audio waveforms~\cite{donahue2019wavegan}, and even new clothing for fashion~\cite{gansclothing}. 

A GAN's fundamental form comprises two neural networks: a generator and a discriminator. The generator network samples random vectors from a high dimensional prior distribution (e.g. Gaussian) as input, and transforms this input vector into a generated image through a series of non-linear upsampling operations. This prior distribution is typically called the latent space for the GAN, and encodes semantic information (e.g. type of object) once the GAN has been trained, although this information is not fully interpretable or explicit. The discriminator network is presented the generator's output image as well as a real image from the training dataset, and then tasked with labeling generated images and real dataset images as either fake or real. In summary, a GAN trains its generator to produce more plausible images similar to the training dataset, and the discriminator to be the critic and force the generator to improve its performance in order to fool the discriminator. This process is known as adversarial training. Formally, the GAN loss function defines a minimax game which encourages the discriminator to assign correct labels and the generator to trick the discriminator into assigning the incorrect labels, and the solution to this minimax game minimizes the Jensen-Shannon divergence between the generated and training image distributions~\cite{goodfellow_GANs}.

This adversarial training can be difficult in practice due to convergence issues. There have been a myriad of suggested improvements for training GANs~\cite{karras2017progressive, salimans2016improved, karras2018style}. One of the most effective methods is the Wasserstein-GAN (WGAN), which replaces the original GAN loss function that minimized the Jensen-Shannon divergence between generated and training distributions, with one that minimizes the more stable Wasserstein-1 distance~\cite{wassGAN,gulrajani2017improved}. We utilize this architecture extensively in our experiments.

\textbf{Scene control.} One major drawback of GANs is the inability to control the scene content in generated images, such as the position and appearance of objects. The typical GAN framework samples a random noise input vector and produces an image that appears to belong to the training set statistically, but the user has little to no control to edit the contents of this image. There have been some attempts to remedy this limitation in the literature, but it usually offers coarse control over the latent space~\cite{chen2016infogan, bau2018gan, mirza2014conditional, karras2018style}. Image-to-image translation~\cite{Zhu_2017, Isola_2017} maps images from one domain to another, e.g. changing black-and-white images to color or daytime landscapes to nighttime landscapes. Style transfer uses the layers of convolutional neural networks to transfer the ``style" of images from one domain to another, such as a regular image to be ``painted" in the style of Monet or Van Gogh~\cite{Johnson_2016}. Our contributions adopt methods from image-to-image translation and style transfer literature as we learn the mapping from rendered optical images to a training dataset of SAS images.

\textbf{GANs for Sonar.} While GANs have been used to generate sonar images of bare seafloor, their potential for generating physically realistic targets on the seafloor is relatively unexplored. Chen et al.~\cite{chen2016deep_neural_networks} train GANs on seafloor images and demonstrate that generated images appear both realistic and novel. Style transfer was used to place targets on the seafloor, however this method did not always infer shadow geometries or glint effects. Lee et al.~\cite{Lee:2018a} also use style transfer to color acoustically rendered images with the global style statistics of real sonar images. Sung et al.~\cite{Sung:2019a} use a conditional GAN~\cite{Isola_2017} for paired image-to-image translation of optically captured objects to insonified objects for forward-looking sonar applications. In a recent study that is closest to our approach~\cite{Karjalainen:2019a}, an unpaired image-to-image translation GAN generates targets on the seafloor by using a ray-tracer to insert rendered images onto real seafloors. Shadow geometry and glint lines are then calculated explicitly using elevation maps. In contrast, our approach refines the entire target and seabed jointly, and does not require prior information of the seabed, or the calculation of shadow geometries and glint lines since it infers these properties from the training data. 

\subsection{Simulated SAS Images}
\label{sec:acousticsim}
Modeling the time series of the acoustic field scattered from a scene can be approached through a variety of techniques. The exact calculation of the field scattered from a boundary requires solving the Helmholtz-Kirchhoff integral equation over a set of specified boundary conditions. Analytic solutions to this integral equation exist for a small number of very simplistic target shapes. For more complex scenarios, the solution may be calculated to arbitrary accuracy through numerical techniques such as the boundary element or finite element methods. While these methods are exact, they come with a significant computational burden and are very difficult to scale up to large scenes and/or high frequencies. Frequently, researchers developing a high-frequency SAS model will resort to approximate methods to accelerate the model computation.

A common type of approximate model discretizes the surface into individual scattering elements, independently calculates the fields scattered from each discrete element, and integrates these individually scattered fields. Under this type of model, multiple scattering is ignored as well as elastic scattering mechanisms. There are several existing models that have approached this problem by discretizing the scene into a triangular mesh and applying the Kirchhoff approximation to these individual facets \cite{sammelman2000computer,Sammelmann:2001a,Sammelmann:2003a,DeTheije:2006a,Hunter:2006a,Abawi:2016a}. Point scattering and point-based scattering models have also proven capable of generating representative time series for a simulated scene \cite{Chen:2009b,Brown:2017b}. While these approximate techniques are relatively fast, they still have a high computational burden. For a simulation of a high-resolution SAS scene the simulation of a single image can require hours of computation.

The simulation community is attempting to address this drawback through several different methods. Direct generation of magnitude imagery is being approached through generative adversarial networks \cite{chen2016deep_neural_networks,Karjalainen:2019a,Sung:2019a}. Additionally, recent research programs have investigated insertion of realistic targets into preexisting real datasets~\cite{Mignotte:2009a}. Each of these approaches attempts to generate larger quantities of either fully synthetic or hybrid data than are currently available from fielded sensors. These larger datasets are being developed with the goals of using them for machine learning applications.

% % %

\begin{figure*}[h]
                \centering
                \begin{tabulary}{\textwidth}{C C}                                                             
                                \includegraphics[height=4.7cm]{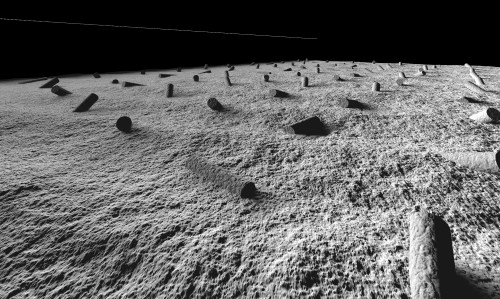} &
                                \includegraphics[height=4.7cm]{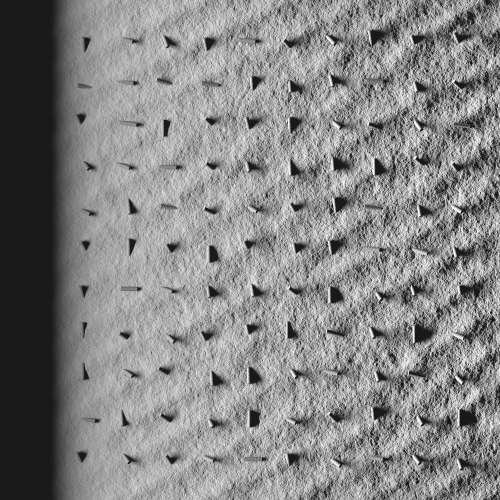} \\
                                (a) & (b) \\
                                \includegraphics[height=4.7cm]{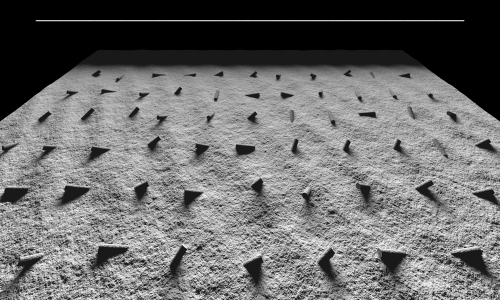} &
                                \includegraphics[height=4.7cm]{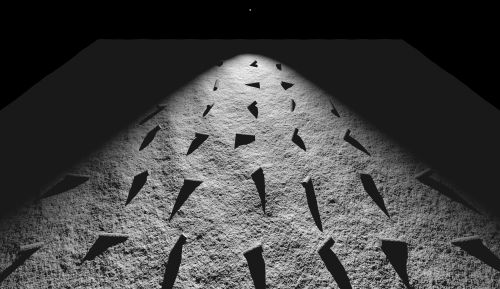}\\
                    (c) & (d)
                \end{tabulary}
                \caption{Artificial target field rendered four different ways to demonstrate the scene geometry and light source beam patterns. The light sources are denoted as point spheres in the scenes. In (a) and (c), the high density of light sources presents as a solid line. (a) Viewpoint rendered from just above the seafloor at range. (b) Rendered scene using the SAS imaging geometry for viewpoint. The scene represents an area of 60m x 60m with cylinders of random orientation and burial depth spaced 5m apart. (c) Array of light sources viewed at max range. (d) Single light source viewed at max range where the conical shaped beam pattern used in the simulations is visible. }
                \label{fig:artifical_target_field_renders}
\end{figure*}

\section{Approach}
\label{sec:approach}

Our approach leverages both model-driven and data-driven methods to simulate SAS images. In this section, we discuss the components of our proposed pipeline including the choice of optical rendering over acoustic rendering and the use of generative adversarial networks. We find this combination to be complementary in their benefits and drawbacks, and validate this in our experimental results in Section~\ref{sec:results}.

We use the term \textit{physical realism} for SAS data that satisfy the acoustic physics and have been beamformed from raw data. Note that in this paper, we do not claim physical realism as we are not modeling any acoustical properties explicitly in our machine learning. We use the term \textit{SAS image realism} or \textit{SAS realism} for short, for images that qualitatively look similar to real SAS images for the dynamic-range compressed magnitude image. The goal of this paper is to generate artificial images that have high SAS realism. 

\subsection{Optical Rendering of SAS Images using POV-Ray}
As noted in Section~\ref{sec:acousticsim}, there exist several simulators to model the acoustic field for SAS image formation, but these models are either computationally expensive (hours to render an image) or provide limited control over the geometry of the models, seafloor, environment, and acoustic sources during rendering. For this reason, we decided to render SAS images using an optical renderer, modeling acoustic sources as optical sources and simulating ray tracing to form the image. Optical renderers solve the Rendering equation~\cite{Kajiya:1986:RE:15886.15902} using Monte Carlo integration, and have been developed to be reasonably fast (order of seconds to minutes for an image) with high control over scene content creation. We acknowledge that to acquire these benefits, we sacrifice both physical realism by using an optical renderer as opposed to an acoustic renderer, as well as SAS realism since our output rendered images do not look like real SAS images. While the first limitation cannot be overcome, achieving better SAS realism can be achieved by our data-driven methods explained in the next subsection.  

We utilize the Persistence of Vision Ray Tracer, simply known as POV-Ray~\cite{povray}, a ray tracing software which renders a scene
in the optical wavelength regime. It ingests as input a text file describing the scene (including light sources, camera
position, and objects) and outputs an image as seen by the camera through its associated parameters. POV-Ray is capable
of rendering non-trivial scattering physics such as reflection, refraction, and texture. We use POV-Ray to generate a grayscale optical image with similar observation properties as a SAS image operating at high frequency (specifically $\frac{f_{max}}{f_{min}} \leq 1.5$)\cite{kohntopp2018shape}. 

We accomplish this by creating a scene consisting of four
items: 1) Array of light sources, 2) seafloor height map~\cite{johnson2009synthetic, tang2009simulating, Johnson:2018a}, 3) objects with associated location \& orientation parame-
ters, 4) a camera descriptor specifying an overhead view and
orthographic projection. The items are placed in the scene in correspondence to a typical SAS geometry with the sonar altitude operating at one-tenth the maximum imaging range. The array of light sources is necessary to mimic the illumination given by the synthetic aperture geometry. The camera is positioned directly above the center of the seafloor scene consistent with the viewpoint
of synthetic aperture imagery. Figure~\ref{fig:artifical_target_field_renders} shows several views of the simulated scene for reference, as well as the scene rendered from the viewpoint mimicking SAS imagery. Notice how the SAS realism is low for these images, including the absence of effects such as speckle. This motivated us to explore data-driven methods for SAS image generation.

%\begin{figure*}
%	\centering
%		\includegraphics[width=0.3\textwidth]{\figPath POV-Ray/geometry_plots_for_oceans_paper.png} 
%		\includegraphics[width=0.3\textwidth]{\figPath POV-Ray/geometry_plots_for_oceans_paper_single_beam.png} 
%		\includegraphics[width=0.3\textwidth]{\figPath POV-Ray/simulated_sas_image.png} 

%	\caption{Artificial target field rendered three different ways to demonstrate the scene geometry and light source beam patterns. The light sources are denoted as point spheres in the scenes. From left to right: (1) array of light sources viewed at max range (2) single light source, viewed at max range, conical beam pattern assumed (3)  Rendered scene using the SAS imaging geometry for viewpoint. The scene represents an area of 60m x 60m with cylinders of random orientation and burial depth spaced 5m apart.}
%	\label{fig:artificial_target_field_renders}
%\end{figure*}

%\begin{figure*}
%\includegraphics[width=0.3\textwidth]{\figPath POV-Ray/geometry_plots_for_oceans_paper.png}
%\includegraphics[width=0.3\textwidth]{\figPath POV-Ray/geometry_plots_for_oceans_paper_single_beam.png}
%\includegraphics[width=0.3\textwidth]{\figPath POV-Ray/simulated_sas_image.png}
%\caption{Rotated view of cylinder renders from POV-Ray.}
%\label{fig:povRay}
%\end{figure*}

\subsection{SAS Image Rendering using GANs}
To achieve SAS realism for our generated images, we turn to recent advances in machine learning, namely generative adversarial networks (GANs). As mentioned in Section~\ref{sec:GANrelated}, GANs allow for high photorealism for optical images, although they lack control over scene content due to their probabilistic nature of sampling from the latent space. To test the effectiveness of GANs for SAS image generation, we use the common DCGAN architecture~\cite{radford2015unsupervised}  trained on our SAS dataset using the Wasserstein with gradient penalty value function \cite{gulrajani2017improved}. Our DCGAN learns to generate SAS images of objects on the seafloor with the visual qualities of our training dataset. 

In Figure~\ref{fig:dcgan}, we show the results of training DCGAN on real SAS images. We observe that generated images contain only objects with positions and rotations found in the training set. We note that more advanced GAN architectures that employ progressive growing strategies ~\cite{karras2017progressive}, ~\cite{karras2018style} can further improve SAS realism, but there are no practical ways to control the scene content such as location of the objects in the scene or the shadows.

\begin{figure}
\centering
\includegraphics[width=0.4\columnwidth]{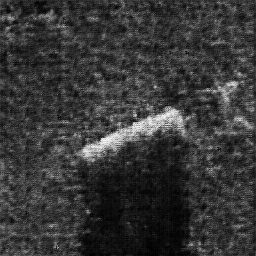}
\includegraphics[width=0.4\columnwidth]{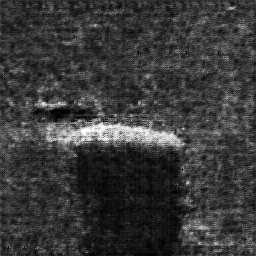}
\caption{Two generated examples from DCGAN. We noted that DCGAN mimicked the training dataset by only generating objects centered in the tile.}
\label{fig:dcgan}
\end{figure}

\subsection{Hybrid POV-RAY to WGAN Architecture}
Our main contribution in this paper is the proposed novel pipeline that synergistically combines POV-Ray and GANs in order to fabricate SAS images.  In Figure~\ref{fig:pipeline}, we detail our proposed pipeline where POV-Ray generates synthetic SAS images, and then a GAN improves these SAS images to reflect the statistics and visual quality of real SAS images. This solution allows for fast rendering of SAS images, with high levels of SAS realism, while allowing the user to control the geometry of the model, its location and orientation along the seafloor, and the placement of the sources in the scene. In other words, our GAN is conditioned (similar to a conditional GAN~\cite{mirza2014conditional}, ~\cite{Isola_2017}) on the POV-Ray image and preserves scene parameters such as target geometry while updating the SAS image realism of the background and scattering. 

One of the challenges of our application domain as compared to rendering natural images is the lack of data to train our data-driven models. To overcome this, we introduce a feature extraction step using an autoencoder within our pipeline that effectively allows semantic features such as target geometry and orientation to flow from the POV-Ray rendered image to the final output.

\textbf{Pipeline.} We mathematically formulate our pipeline as follows. Let $\rho_{p}$ be the distribution of POV-Ray rendered SAS images, and $\rho_{r}$ the distribution of real SAS images. We seek a function $G: \rho_{p}\rightarrow\hat{\rho}_{p}$ that transforms the rendered image distribution such that the statistical distance between $\hat{\rho}_{p}$ and $\rho_{r}$ is minimized. This in effect enforces SAS image realism in our pipeline. Since we do not know our image distributions $\rho$ explicitly, we lack a way to directly find an optimal function for $G$. However, using only the rendered images and real image samples we have available, the Wasserstein GAN training routine allows us to indirectly minimize the distance between these two distributions by finding a $G$ that satisfies the Kantorovich-Rubinstein duality~\cite{wassGAN}.

\begin{figure*}\centering
\includegraphics[width=\textwidth]{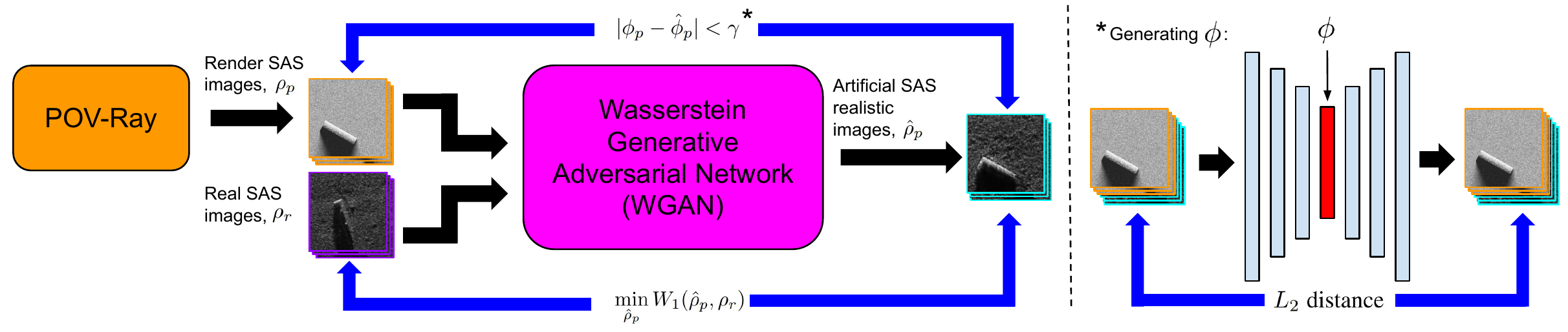}
\caption{Our proposed POV-Ray to GAN pipeline for generating artificial SAS images. The left side of the dashed line shows our image generation pipeline. The top equation shows our constraint for maintaining scene descriptors between rendered and transformed images. The bottom equation shows our minimization of the Wasserstein-1 distance between $\rho_{r}$ and $\hat{\rho}_{p}$. The right side of the dashed line illustrates generating $\phi$ from our autoencoder that was trained on $\rho_{r}$.}
\label{fig:pipeline}
\end{figure*}

To ensure our output images still preserve the target geometry and other scene parameters, we define $\phi$ to be a feature extractor that maintains $|\phi_{p} - \hat{\phi}_{p} | < \gamma$, where $\gamma$ is a tuneable hyperparameter, and  $\phi_{p}, \hat{\phi}_{p} $ represent the high-level scene descriptors (object position, seafloor style, etc.) of our rendered and transformed images respectively. To enforce the constraint for $\phi$, we extract feature vectors from a convolutional autoencoder (AE) trained on $\rho_{r}$. Our convolutional autoencoder is a neural network tasked with reducing each high dimensional image in $\rho_{r}$ into a smaller dimensional column vector. To most effectively compress the dimensionality of the $\rho_{r}$ images, our convolutional AE learns a set of feature filters that capture the salient properties of these images. We constrain the AE to reducing images into a dimension that contains enough information to preserve the global properties of the image, such as scene descriptors like object placement and shadows, but allows higher order effects necessary for SAS realism to be determined by the GAN. Since our autoencoder learns features from ${\rho}_{r}$, we are biased toward preserving ``realistic" scene parameters between $\hat{\rho}_{p}$ and $\rho_{p}$. 

\textbf{Network Architecture.} 
We adopt a similar generator architecture to~\cite{Johnson_2016} of a series of downsampling convolution layers, a series of residual layers, and a series of upsampling+convolutional layers. We use the discriminator network from DCGAN which consists of a series of downsampling convolutional layers. We train over a value function that minimizes the Wasserstein-1 distance between $\rho_{r}$ and $\hat{\rho}_{p}$. The value function,  

\begin{equation}
	\begin{aligned}
		\min_{G}\max_{D \in \mathcal{D}} & \mathop{\EX}_{x \sim \rho_{r}}[D(x)] -  \mathop{\EX}_{\hat{x} \sim \hat{\rho}_{p}}[D(\hat{x})] - \\
							   & \lambda \mathop{\EX}_{\hat{x} \sim \rho_{\hat{x}}}[(||\nabla_{\hat{x}}D(\hat{x})||_{2} - 1)^2],
	\end{aligned}
	\label{val}
\end{equation}
is created from the Kantorovich-Rubinstein duality as shown in ~\cite{wassGAN, gulrajani2017improved}.

The Wasserstein-1 distance between $\rho_{r}$ and $\hat{\rho}_{p}$ is minimized when we find a $D$ network, which is encouraged to be 1-Lipschitz smooth by the gradient penalty term, that maximizes this value function. Here, $\hat{x} \sim \hat{\rho}_{p}$  are generated images from the distribution created through $G: \rho_{p}\rightarrow\hat{\rho}_{p}$, and $\hat{x} \sim \rho_{\hat{x}}$ are randomly sampled generated samples for calculating the gradient norm of $D$.

%\begin{equation}
%W_{2}(\rho_{A}, \rho_{B}) = ||\phi_{A} - \phi_{B}||_{2}^{2} + \Tr(C_{A} + C_{B} - 2(C_{B}^{\frac{1}{2}}C_{A}C_{B}^{\frac{1}{2}})^{\frac{1}{2}})
%\end{equation}

In order to obtain a set of filters for extracting $\phi$, we train our autoencoder using a series of four downsampling covolutional layers, each followed by ReLU activations and max-pooling, to compress our 256 $\times$ 256 input images into a 1024-dimensional column vector, $\phi$, and then four nearest-neighbor interpolated upsamplings followed by convolutions to reconstruct the image from the compressed vector. After training, we disregard the upsampling layers and use the downsampling layers to generate  a 1024-dimensional $\phi$ vector for a given image. We train our autoencoder over an $L_{2}$ distance between input and reconstructed images. 

% \begin{equation}
%     L_{2} \; \text{distance}
% \end{equation}

% % %
% Experimental Results
\section{Experimental Results}
\label{sec:results}

We present experimental work to illustrate the potential quality of SAS GAN with respect to genuine sonar image data. Our hybrid pipeline produces fabricated SAS images with realistic SAS-image characteristics. We also show control over the target and source geometry and location in the scene due to POV-Ray. To analyze our results, we employ quantitative metrics to relate fabricated SAS image quality fidelity to genuine SAS data, and discuss challenges of evaluating results for data-driven methods for this application domain. 

% We validate that our hybrid pipeline can produce fabricated SAS images with realistic SAS-image characteristics borne from the injection of POV-Ray into the generation process in a manner that a typical GAN without such preprocessing cannot. We also show control over the target and source geometry and location in the scene due to POV-Ray. To analyze our results, we employ quantitative metrics to relate fabricated SAS image quality fidelity to genuine SAS data, and discuss challenges of evaluating results for data-driven methods for this application domain. 

\subsection{Implementation Details}

\textbf{SAS Data.} Our genuine SAS data used in this section comes from the MUSCLE autonomous underwater vehicle courtesy of the NATO Centre for Marine Research \& Experimentation. This system has a high-frequency SAS developed by Thales with a center frequency of 300 kHz and bandwidth of 60 kHz. The imagery from MUSCLE has approximate resolutions of 2.5 cm and 1.25 cm in the along-track and across-track directions, respectively, and can reach out to 150m in range~\cite{williams2019novel}. We note that we use only 560 images containing cylindrical objects in this dataset, which is very data-starved for deep learning applications. The images contained a diversity in both background types and target orientation, though we note that all the targets were centered in their images as this will be important later. 

\textbf{POV-Ray specifications.} We render cylindrically shaped targets with a size approximated to the targets in the dataset. These targets are rendered with a rough surface on a seabed of Gaussian distributed pixel values. The noise on the targets and backgrounds encourages stochasticity in generated samples. 

\textbf{Training details.} 
We train our SAS GAN pipeline on 2 Titan-X GPUs for 24 hours with 560 real images and 850 POV-Ray images. All images are sized 256 $\times$ 256 pixels. We optimize the Wasserstein with gradient penalty value function shown in Equation~\ref{val} (with gradient penalty set to 10) using the Adam optimizer with a learning rate of 0.001 and batch size of 4.

\subsection{Qualitative Results}

\begin{figure*}
\includegraphics[width=\textwidth]{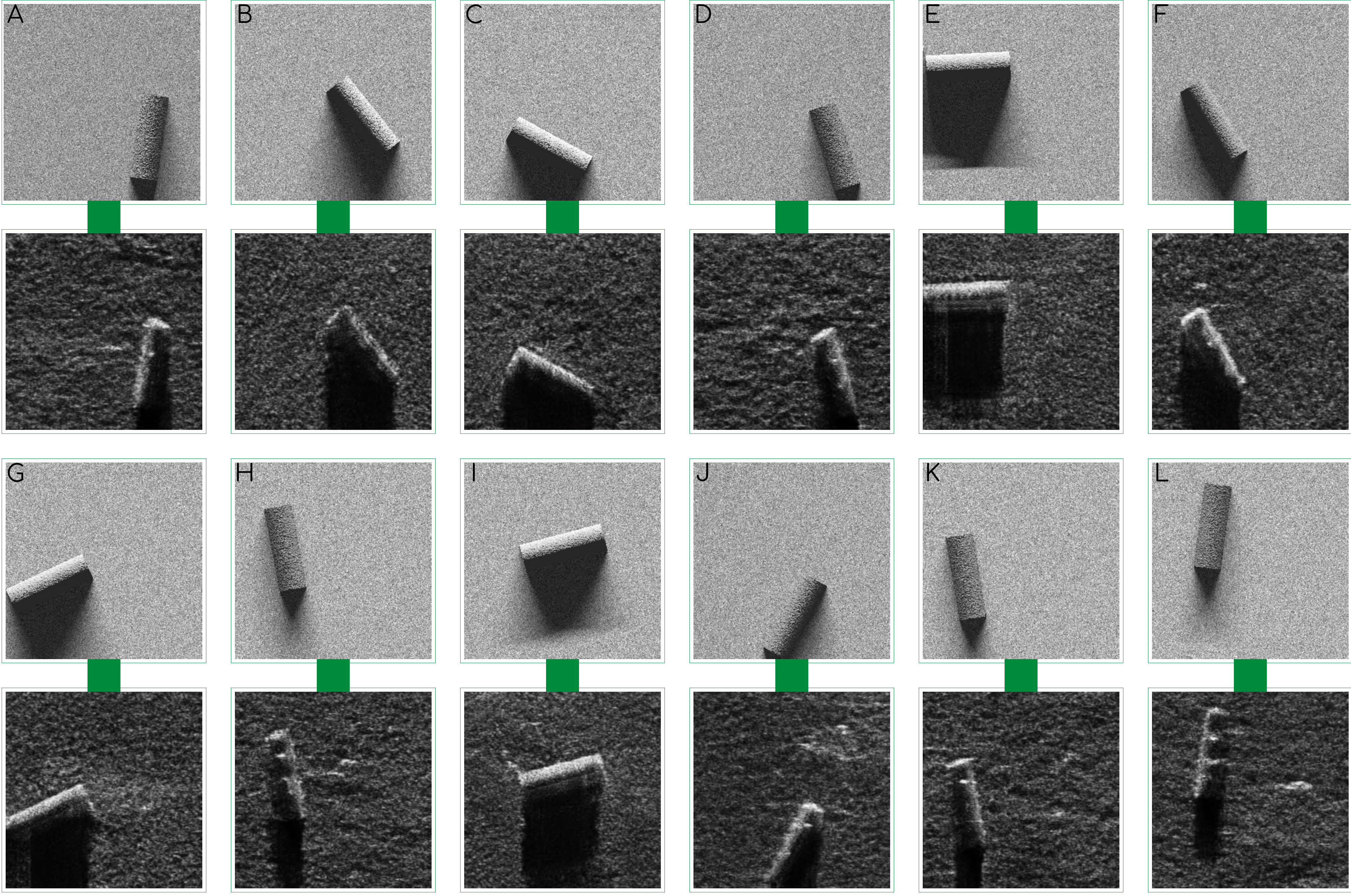}
\caption{Example SAS GAN images. The POV-Ray (lettered) inputs are connected by green lines to their associated SAS GAN images.}
\label{fig:generatedExamples}
\end{figure*}

Figure~\ref{fig:generatedExamples} displays several POV-Ray input images and the associated SAS GAN output images. SAS GAN captures the image resolution and blurring common to the SAS images by dramatically altering the given POV-Ray rendered inputs. This shows the network is performing a style transfer from POV-Ray images to SAS images while still retaining the target in the image. SAS GAN is acting as intended: it is given a controlled target input and produces a realistic-looking output with a target in the same position.

% An immediate impression is that SAS GAN is able to capture the type of resolution and blurring common to SAS images by dramatically altering the given POV-Ray render. Indeed, the type of blurring and highlighting seen in each case shows that the network is performing a style transfer from the pristine POV-Ray environment to SAS while still retaining the target in the image. In that way, SAS GAN is acting as intended: it is given a controlled target input and is able to yield a realistic-looking output with a target in the same position.

Note that SAS GAN learns to alter the shadow pattern of the input render. In each case of Figure~\ref{fig:generatedExamples}, the given POV-Ray shadow is altered to match shadows more characteristic of SAS images, namely by becoming longer and darker. This better aligns with shadows that would be formed from targets at these type of distances away from the vehicle.

%what would be found at the type of distances these targets would be away from the vehicle. As well, many of the shadows encroach inward on the target, like with pairs N and O. This behavior again demonstrates how SAS GAN tries to alter the perspective of its fabrications to match the perspective of the MUSCLE system which, in this case, relates to the grazing angle view.

Along with adjusting the shadow, SAS GAN has learned to generate fine-grained details visible on targets. Several targets exhibit strong circular patterns on top of the cylinder. These details are learned from the MUSCLE system and relate to realistic target shapes. SAS GAN produces these realistic targets \emph{without instruction} (i.e. a regularization technique that encourages such behavior). SAS GAN is not just learning a simple style transfer of where to place shadows or background texture; it is actively figuring out what a target is. We argue that our qualitative results are compelling and have high SAS realism; this was informally confirmed by several SAS researchers and engineers who viewed the images.

% capture minute details and applies these minute details them to cases where the target is at a given position in order to make these attributes visible, \emph{without instruction} (i.e. a regularization technique that encourages such behavior). This means that SAS GAN is not just learning a simple style transfer of where to place shadows; it is actively figuring out what a target is. We argue that our qualitative results are compelling and have high SAS realism; this was informally confirmed by several researchers and engineers working on SAS that we showed these images to. 

%Again referring to Figure~\ref{fig:generatedExamples}, examples M and P (among others) show a strong circular pattern on top of the cylinder. These details are learned from the MUSCLE system and relate to realistic target shapes. This is all to say that the SAS GAN is able to capture minute details and apply them to cases where the target is at a given position in order to make these attributes visible, \emph{without instruction} (i.e. a regularization technique that encourages such behavior). This means that SAS GAN is not just learning a simple style transfer of where to place shadows; it is actively figuring out what a target is. We argue that our qualitative results are compelling and have high SAS realism; this was informally confirmed by several researchers and engineers working on SAS that we showed these images to. 

\textbf{Novelty of generated images.} One of the difficulties of training in a data-starved regime is that the GAN can learn to identically copy the training set and reproduce those images at test time. This means that the network has not learned to generalize across the image dataset, or not effectively learned the input distribution to sample novel points (images) from that distribution. To ensure that our network is indeed learning the input distribution, we checked the nearest neighbor in both image space ($\ell_2$ loss) as well as feature space $\phi$ to ensure no image was copied. 

% This would show that the network had not learned to generalize across the image dataset and effectively learn the input distribution to sample new points (images) from that distribution. To ensure that our network is indeed learning the input distribution, we checked the nearest neighbor in both image space ($\ell_2$ loss) as well as feature space $\phi$ to ensure no image was copied. 

Figure~\ref{fig:nearestNeighbor} shows the results of our nearest neighbor experiment. We confirm that SAS GAN is not simply copying the MUSCLE data as the closest real images are markedly different. The nearest neighbor MUSCLE images returned are also cylinder images, however they do not always align with the target geometry present in SAS GAN. Sometimes the nearest neighbor images had unusual artifacts (C) or repetitive images (D-F), potentially caused by measuring distance in feature space $\phi$ which can overlook such visual inconsistencies.

% Encouragingly, the nearest neighbors returned 

% a stylistic commonality exists between each of the SAS GAN outputs and real images and, in each case, the nearest neighbor to each fabrication is a clean image of a cylinder. Further, we see that the SAS GAN is not simply copying the MUSCLE data as the closest real images appear to have distinctly different aspects. 

% What is unexpected are the bizarre nearest neighbors like in example C and the repetitious MUSCLE images that appear for examples D, E, and F. For those, we look deeper into the feature distribution of the SAS GAN.

 \begin{figure*}
\centering
\includegraphics[width=\textwidth]{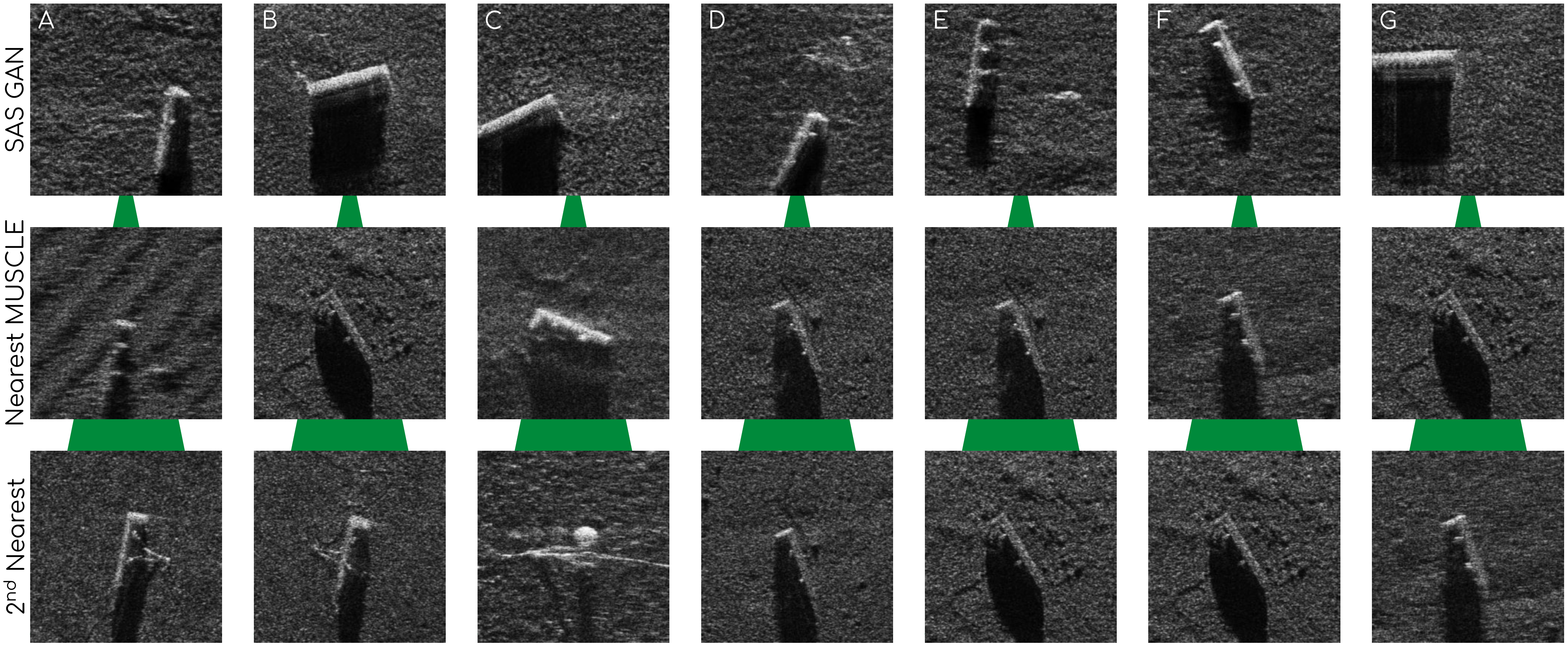}
\caption{Nearest neighbor results for a selection of SAS GAN fabrications (top row) and genuine target examples from MUSCLE (bottom three rows) using the autoencoder distance.}
\label{fig:nearestNeighbor}
\end{figure*}

%left off here

% %
% Translation and Rotation
\subsection{Translation, Rotation, and Background Diversity} One of our stated goals was to allow target control over geometry and location in the image. This is important since the MUSCLE training data did not have all possible orientations of the cylinder and all targets were explicitly centered in the images. To generate novel and diverse images, our hybrid pipeline must preserve a POV-Ray input target's location and orientation while increasing the SAS realism of the output. 

Figure~\ref{fig:translationrotation}(a-b) shows the results of such translation and rotation experiments respectively, in which the target location in the POV-Ray images was varied. SAS GAN is able to sustain realistic target scenarios throughout these transformations. The translation experiment shows that SAS GAN preserves the target's diagonal movement with little change in the target's appearance. Similarly, the rotational effects on the shadows appear in line with what we would expect. It is interesting to note the case where the cylinder is viewed at endfire; the face nearest the vehicle scatters strongly while the rest of the target blends into the background. This suggests that, while SAS GAN is able to impressively model many different views of the cylinder, this exact case lies on the edge of its capabilities. Either more training is required to solve this or a more nuanced approach is needed. Regardless, the overall performance in both translation and rotation is encouraging.

\begin{figure*}[t!]
    \centering
    \begin{subfigure}[t]{0.48\textwidth}
        \includegraphics[width=\textwidth]{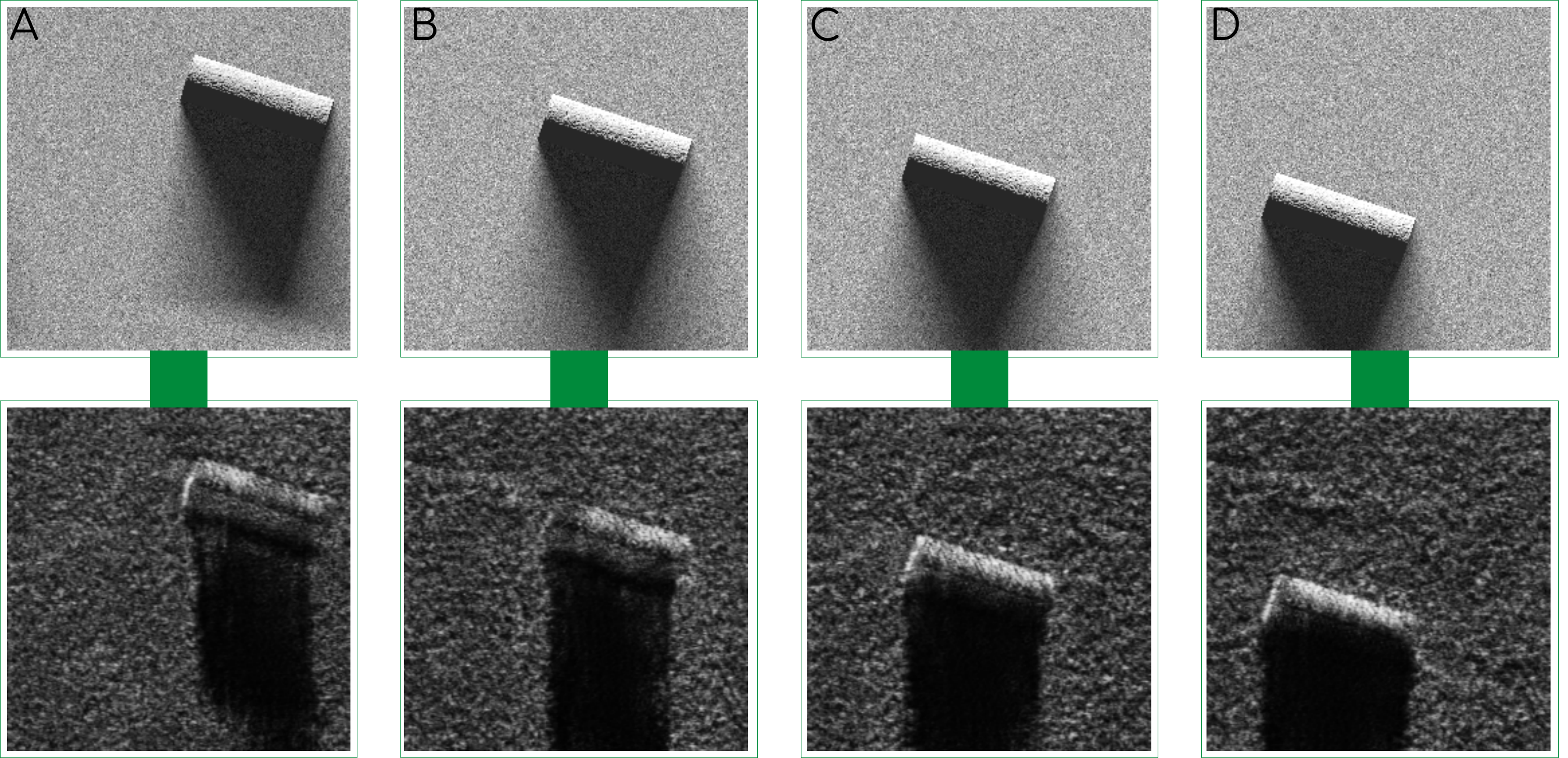}
        \caption{Translated POV-Ray inputs}
    \end{subfigure}
    ~ 
    \begin{subfigure}[t]{0.48\textwidth}
        \includegraphics[width=\textwidth]{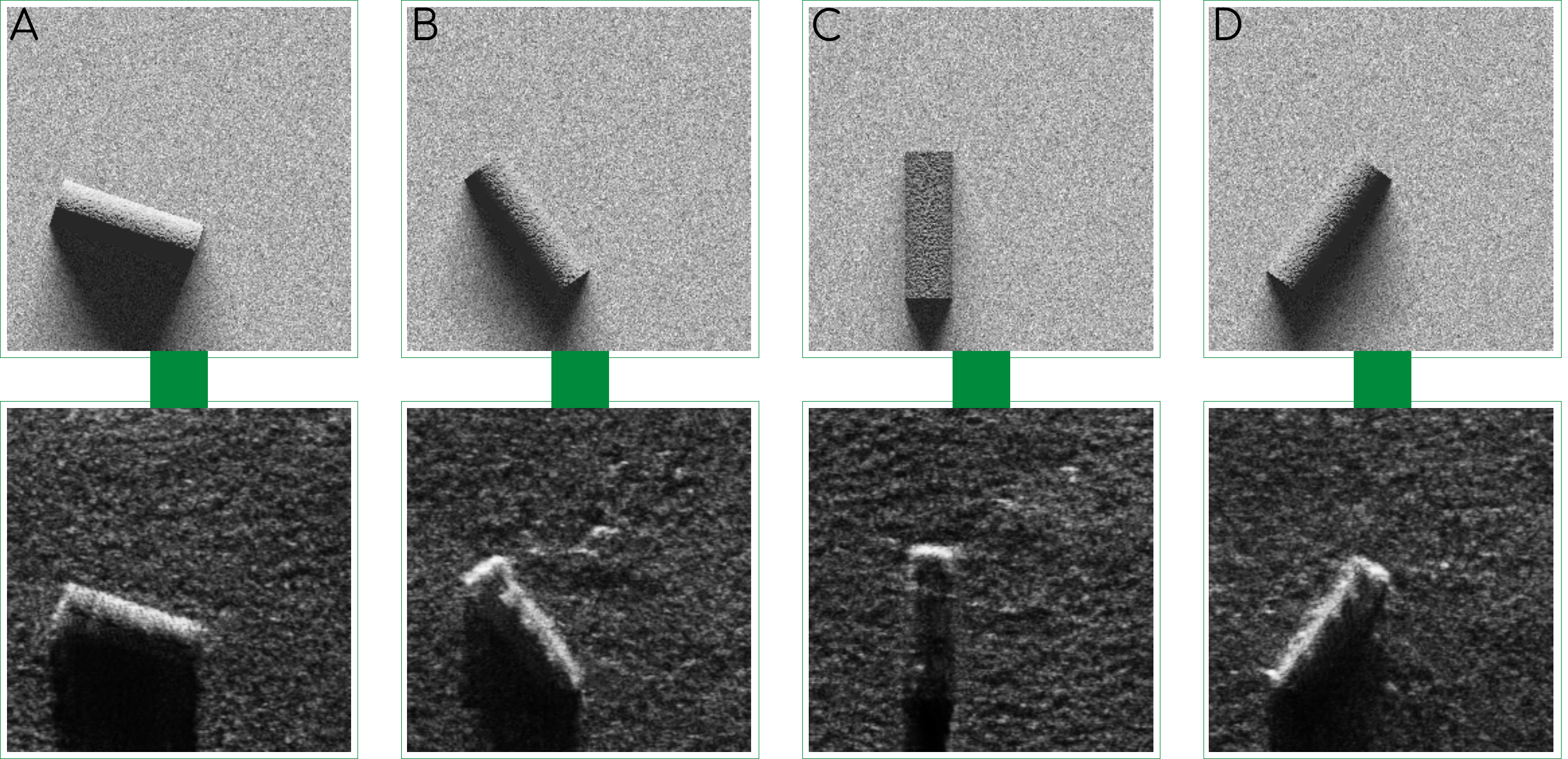}
        \caption{Rotated POV-Ray inputs.}
    \end{subfigure}
    \caption{Example SAS GAN images for translated (a) and rotated (b) POV-Ray inputs.}
    \label{fig:translationrotation}
\end{figure*}

%  \begin{figure*}
%  \centering
%  \includegraphics[width=0.48\textwidth]{Translation_crop.pdf}
%\includegraphics[width=0.48\textwidth]{Rotation_crop.pdf}

%  \caption{Example SAS GAN images for a translated POV-Ray input. \textcolor{red}{make subfigures, make sure labeled in text correctly}}\label{fig:translationrotation}
%  \end{figure*}
 
%  \begin{figure*}
%  \includegraphics[width=\textwidth]{Translation.pdf}
%  \caption{Example SAS GAN images for a translated POV-Ray input.}\label{fig:translation}
%  \end{figure*}
 
%  \begin{figure*}
%  \includegraphics[width=\textwidth]{Rotation.pdf}
%  \caption{Example SAS GAN images for a rotated POV-Ray input.}\label{fig:rotation}
%  \end{figure*}
 
Another key aspect to SAS GAN's output is the background. The POV-Ray render has the target with shadow encompassed by speckle noise so we do not impart some control over the seafloor like we do the cylinder. This results in the types of textures illustrated in Figure~\ref{fig:seafloors}. Here we see a few different seafloor types. While the differences can be subtle, like in examples C, D, and E, they are distinct from one another. SAS GAN also learns to put an occasional clutter item, like in G, despite no instruction.

\begin{figure*}
\includegraphics[width=\textwidth]{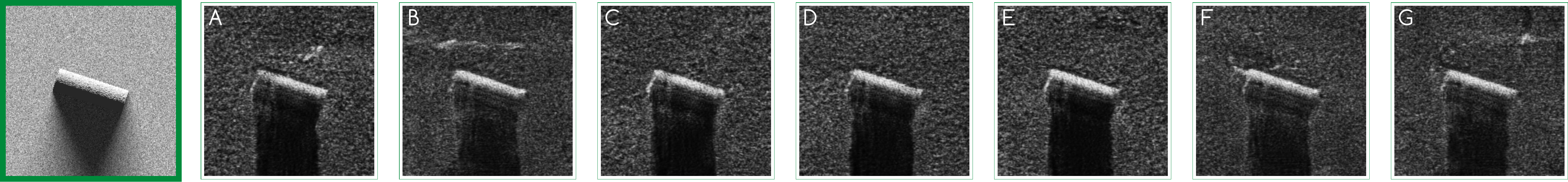}
\caption{Example seafloor generations for a given target. The far left image is an example POV-Ray input and the rest are SAS GAN fabrications. Note that the only difference to the SAS GAN's input to generate these different backgrounds was the noise outlining the POV-Ray target.}
\label{fig:seafloors}
\end{figure*}

\subsection{Quantitative Results}
One of the challenges of evaluating data-driven methods for image generation is the lack of suitable quantitative metrics to evaluate image quality. This difficulty has been noted in multiple domains, spurring work on reference and reference-free visual quality assessment~\cite{lin2011perceptual}. However, we still evaluate some common metrics for GAN performance to elucidate some insights into the network's behavior.

\begin{table}
\centering

\begin{tabular}{|c|c|}
\hline
\textbf{MUSCLE$\to$MUSCLE} & $2.324 \pm .038$ \\\hline
\textbf{SAS GAN$\to$MUSCLE} &  $2.934 \pm .046$ \\\hline
\textbf{DCGAN$\to$MUSCLE} & $2.289 \pm  .035$ \\\hline
\textbf{POV-Ray$\to$MUSCLE} & $14.472 \pm .065$ \\\hline
\end{tabular}
\caption{FID Scores. The MUSCLE to MUSCLE score serves as a reference score to give context to the DCGAN, POV-Ray, and SAS GAN scores.}\label{tab:fid}
\label{fid}
\vspace{-0.1cm}
\end{table}

\textbf{FID score.} To quantitatively evaluate GAN output quality, Heusel et al.  proposed the Fr\'echet inception distance (FID) in \cite{heusel2017gans}, also known as the 2-Wasserstein distance. As this metric has been widely adopted in the literature \cite{brock2018large, karras2018style, karras2017progressive}, we use a variant on this score for providing a quantitative metric comparing the similarity of the DCGAN, SAS GAN, and POV-Ray distributions. Given two datasets of images $A$ and $B$, the proposed FID metric uses an intermediate layer from Inception network \cite{Szegedy_2015} pretrained on ImageNet \cite{imagenet_cvpr09} images to capture the average feature vectors, $\phi_{A}, \phi_{B}$, and covariance matrices $C_{A}, C_{B}$ over all images $i \in A, B$. Then if we use these mean vectors and covariance matrices to define $\rho_{A} = \mathcal{N}(\phi_{A}, C_{A})$ and $\rho_{B} = \mathcal{N}(\phi_{B}, C_{B})$ as multivariate Gaussian distributions, the 2-Wasserstein distance between $\rho_{A}$ and $\rho_{B}$ is defined as

\begin{equation}
	\begin{aligned}
		W_{2}(\rho_{A}, \rho_{B}) = &  ||\phi_{A} - \phi_{B}||_{2}^{2} + \\
						      &  \Tr(C_{A} + C_{B} - 2(C_{B}^{\frac{1}{2}}C_{A}C_{B}^{\frac{1}{2}})^{\frac{1}{2}}),
	\end{aligned}
	\label{frechet}
\end{equation}
where $\Tr{}$ is the trace.

\begin{figure*}
\includegraphics[width=\textwidth]{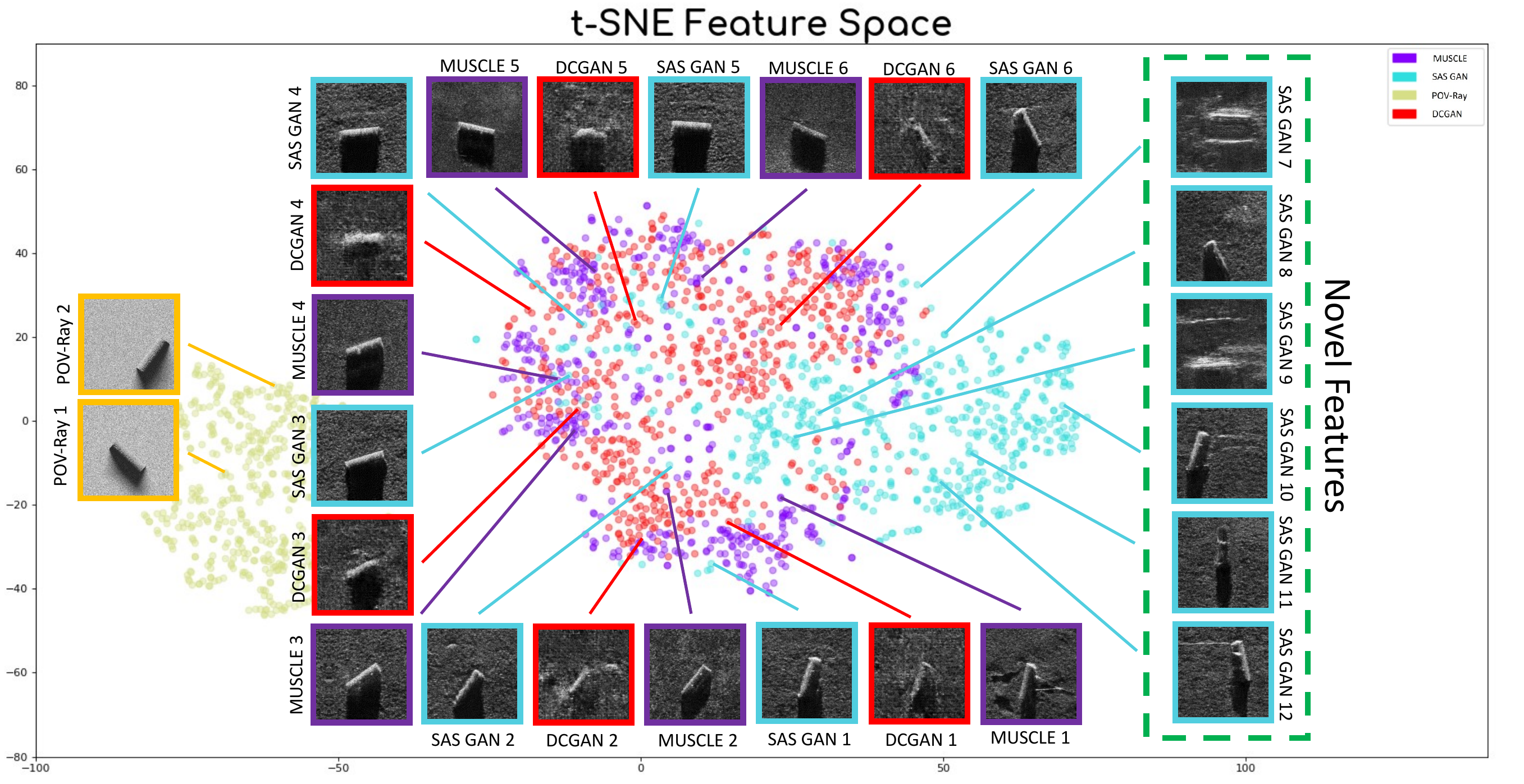}
\caption{t-SNE visualization of the feature space spanned by real MUSCLE images, POV-Ray renders, and fabricated imagery generated by both DCGAN and our SAS GAN. Images are numbered for easy reference.}
\label{fig:tsne}
\end{figure*}

Since the ImageNet feature space is not optimal for capturing SAS image features, we instead utilize the features captured from our autoencoder trained on the dataset of MUSCLE images. The results of these experiments are displayed in Table~\ref{fid}.

%\textcolor{red}{Suren does this make sense? I am trying to explain why the DCGAN to MUSCLE distance is actually lower than the MUSCLE to MUSCLE distance.}

 We first note that the DCGAN to MUSCLE score is actually lower than the baseline reference score of MUSCLE to MUSCLE. This phenomena hints at a limitation of using FID score; it cannot filter out GANs which replicate the training dataset since two identically copied datasets will achieve the best possible FID score of zero. The FID from DCGAN to MUSCLE is slightly lower than the FID from SAS GAN to MUSCLE, and both distances are significantly lower than POV-Ray to MUSCLE. We suspect SAS GAN has a higher FID than DCGAN because SAS GAN has the ability to generate objects in positions and orientations not found in the MUSCLE dataset. In contrast, DCGAN generates objects only in positions and orientations that can be interpolated from images found in the MUSCLE dataset. Thus we hypothesize SAS GAN is actually augmenting the MUSCLE distribution, rather than simply replicating it like DCGAN, and therefore scoring technically worse when its FID score is measured. To obtain more tangible evidence for this hypothesis, we utilized a non-linear dimensionality reduction technique called t-SNE \cite{maaten2008visualizing} to visualize image distribution structure.

\textbf{t-SNE. } t-distributed Stochastic Neighbor Embedding (t-SNE) is a well-established dimensionality reduction technique that summarizes high dimensional data in a digestible Cartesian grid~\cite{maaten2008visualizing}. t-SNE offers insight into the geometries of globally non-linear data, but since its Cartesian mappings are obtained through the non-deterministic minimization of a non-convex function, it is important to consider its results carefully. In Figure~\ref{fig:tsne}, we plot the t-Distributed Stochastic Neighbor Embedding (t-SNE) feature space of both the GAN models, MUSCLE images, and POV-Ray renders when they are passed through the autoencoder features $\phi$.

We will walk the reader through Figure~\ref{fig:tsne}, point out relevant features, and explain how they lend insight into the performance of the different generated images.  Looking at the MUSCLE (purple) dataset, we see they form a ring structure. Notice as one goes from MUSCLE images 1-6 clockwise, the t-SNE position of the image correlates with the target's orientation at the center of a tile. This shows that the autoencoder features semantically disentangled the latent space and learned target orientation as a scene parameter. This is later exploited by our novel $\phi$ loss. Thus, our t-SNE analysis shows that the MUSCLE dataset (1) has a rotational structure in the latent space of the $\phi$ and (2) has no off-center targets.

Looking at the POV-Ray images in yellow, we notice they cluster in t-SNE away from the MUSCLE dataset, and qualitatively the images do not have the same distributional statistics as the more realistic SAS images of SAS GAN and DC GAN (evidenced by the large FID score). However, POV-Ray images do include both centered and off-centered targets.

For DCGAN (red), we show that they also cluster near the MUSCLE dataset and exhibit the same rotation in their latent space. Note however that DCGAN images sampled near the center (e.g. DCGAN 6) are of lower quality because the network is trying to generate novel images in a region for which there are no nearby MUSCLE images.

Finally, we analyze the results of our SAS GAN (light blue). Note SAS GAN images that cluster near the MUSCLE data are qualitatively of the same SAS realism. But, SAS GAN also generates a new cluster of images on the right (e.g. SAS GAN 7-12), showing that the network can generalize and extrapolate beyond the training set to create novel images. However, we do observe a limitation: not all of these sampled images are of high SAS realism. But the new cluster does exhibit off-centered images and varying target orientations, which are not possible for DCGAN and not present in the real MUSCLE dataset. We argue that this new cluster augments the MUSCLE distribution with novel images and explains the discrepancy in FID scores between DCGAN and SAS GAN.

\section{Discussion}

Our hybrid approach coupling optical rendering with GANs yields several interesting discussion points and avenues for future investigation. As validated in our experimental results, we achieve high levels of SAS image realism while enabling control over scene geometry and parameters. Further,  our pipeline achieves fast rendering times at test time: we can render a single SAS image in 250 milliseconds, and thus approximately 900 images in one hour. This is useful for dataset augmentation for data-starved tasks such as ATR. 

There are several limitations to our approach. In Figure~\ref{fig:badCases}, we show some failure cases of our GAN output. These include effects unrealistic shadow geometries and orientations, and unrealistic visual artifacts on the targets. Some of these errors could be mitigated by a larger, well-curated SAS image dataset, which would benefit machine learning approaches. 

\begin{figure}\centering
\includegraphics[width=0.4\columnwidth]{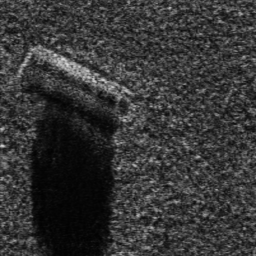}
\includegraphics[width=0.4\columnwidth]{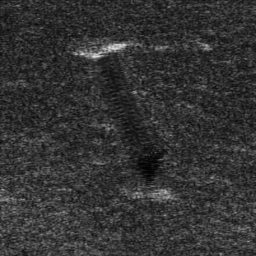}
\caption{Two examples of failure cases: The left image shows a cylindrical object with an unrealistic shadow. The right image shows a poorly generated object with repeating artifacts and unrealistic glint lines.}
\label{fig:badCases}
\vspace{-0.2cm}
\end{figure}

There are several directions for future reasearch. To achieve better physical realism (as opposed to SAS image realism), we can condition the GAN on input seeded from an acoustic simulator such as point-based scattering models~\cite{Brown:2017b}. Further work is needed to quantify the performance of GAN outputs for SAS images. Finally, fully completing the pipeline and evaluating the level of physical realism and control of scene parameters/content generation needed for SAS ATR (similar to~\cite{Karjalainen:2019a}) would help show the advantages of hybrid pipelines.

% % %
% Acknowledgments
\textbf{Acknowledgments:} The authors would like to thank the NATO Centre for Maritime Research \& Experimentation (CMRE) for providing the data used in this work. The collection of the data was funded by the NATO Allied Command Transformation. We also thank Shawn Johnson for providing source code to generate realistic seafloor texture and height maps. 

\bibliographystyle{ieeetr}
{\footnotesize \bibliography{reed2019generating-sub}}

\end{document}